\documentstyle[preprint,aps]{revtex}
\tightenlines
\begin{document}

\title{Curvature and Cosmic Repulsion\footnote{astro-ph/9803135, 
March 11, 1998}}

\author{\normalsize{Philip D. Mannheim} \\
\normalsize{Department of Physics,
University of Connecticut, Storrs, CT 06269} \\ 
\normalsize{mannheim@uconnvm.uconn.edu} \\}

\maketitle

\begin{abstract}
Some recent supernovae studies have extended the distance versus 
velocity Hubble plot to very high redshift, and have revealed the 
apparent presence of a cosmic repulsion. We show that such a 
repulsion has a natural origin within conformal gravity, a fully 
covariant candidate alternative to standard gravity, and that it 
arises from the gravitational field itself via the negative 
spatial curvature which conformal cosmology is required to 
possess. In this same cosmology the flatness, horizon and 
cosmological constant problems are all resolved, and no need is 
found for dark matter.
\end{abstract}
\bigskip

Very recently two independent groups 
\cite{Garnavich1998,Perlmutter1998a,Perlmutter1998b} have reported 
some new high $z$ redshift data which appear to have some 
potentially far reaching implications for 
cosmology. Specifically, by using type Ia supernovae as standard 
candles the two groups were able to extend the Hubble plot of 
distance ($d_L=r_1R^2(t_0)/R(t_1)$) versus velocity 
($z=R(t_0)/R(t_1)-1$) out to redshifts of order 
$z=1$. (Here the source coordinates $(t_1,r_1)$ are related to 
the observer coordinates $(t_0,0)$ via 
$\int_{t_1}^{t_0}dtR(t)^{-1}=
\int_{0}^{r_1}dr(1-kr^2)^{-1/2}$.) The great utility of these new 
data is that they extend way beyond the small $z$ region where
$H(t_0)d_L=z+(1-q(t_0))z^2/2$, to thus enable us to explore
the deceleration parameter $q(t)$ at both low and high $z$. 
Indeed not only do the new data show that $H(t_0)d_L-z$ is 
positive at low $z$ (so that $q(z=0)$ is less than one), the data 
also show that $q(z=1)$ is smaller still. The data thus (very 
qualitatively) suggest that $q(t)$ may actually be an increasing 
function of time, i.e. that the universe may be speeding up.

In order to quantify their results the two groups try to
fit their data to the predictions of the standard Einstein gravity 
cosmology based on the evolution equation
\begin{equation}
\dot{R}^2(t) +kc^2=\dot{R}^2(t)\Omega_{M}(t)
+\dot{R}^2(t)\Omega_{V}(t)
\label{(1)}
\end{equation}
where $\Omega_{M}(t)=8\pi G\rho_{M}(t)/3c^2H^2(t)$ is due to 
ordinary matter and $\Omega_{V}(t)=8\pi G\Lambda/3c^2H^2(t)$ is
due to a cosmological constant $\Lambda$. Specifically, they find 
that the popular standard inflationary universe \cite{Guth1981} 
paradigm ($k=0$, $\Omega_{M}(t)=1$, $\Omega_{V}(t)=0$, 
$\rho_M(t)=A_M/R^3(t)$, $R(t)=t^{2/3}$, 
$q(t)=\Omega_{M}(t)/2=1/2$, 
$H(t_0)d_L=2(1+z)(1-(1+z^2)^{-1/2})$)
also leads to values for $d_L$ which are less than the
observed ones at high $z$, to thus not only exclude the standard
paradigm, but to also show that $q(z=1)$ is quite a bit less than 
$1/2$. The pure vacuum dominated inflationary universe limit 
($k=0$, $\Omega_{M}(t)=0$, $\Omega_{V}(t)=1$, 
$R(t)=exp((8\pi G \Lambda /3c^2)^{1/2}t)$, $q(t)=-1$, 
$H(t_0)d_L=z(1+z)$) leads to values for $d_L$ which are somewhat 
greater than the observed ones at high $z$. The data thus lie 
somewhere between these two extremes (that positive $\Lambda$ 
gives negative $q(t)$ is of course the reason why the 
inflationary universe actually inflates), with $q(t)$ thus lying 
between $-1$ and $+1/2$, and 
with it potentially being negative in the current universe. 
And indeed, the observers find that they are able to fit the data 
in the general inflation case ($\Omega_{M}(t)+\Omega_{V}(t)=1$, 
$\Omega_{V}(t)/\Omega_{M}(t)=(1-2q(t))/2(1+q(t))$, 
$q(t)=-1+(3/2)sech^2\theta(t)$ where $\theta(t)=
3(8\pi G \Lambda/3c^2)^{1/2}t/2+
arcsinh((\Lambda/A_M)^{1/2}R^{3/2}(t=0))$) with best fits 
centering around $\Omega_{V}(t_0)/\Omega_{M}(t_0)=3$, to 
yield a time dependent $q(t)$ with a currently negative 
value of $q(t_0)=-5/8$ (albeit with big 
systematic errors which permit $q(t_0)$ values up to $q(t_0)=0$). 
The groups also explore the non-zero spatial curvature region by 
allowing $\Omega_{M}(t_0)$ and 
$\Omega_{V}(t_0)$ to vary freely and report their results out as 
an allowed region in the $(\Omega_{M}(t_0),\Omega_{V}(t_0))$ space 
which favors large $\Omega_{V}(t_0)$ and small or even zero 
$\Omega_{M}(t_0)$ and is centered around 
$\Omega_{V}(t_0)=\Omega_{M}(t_0)+0.5$. 
In passing they note that the only acceptable fits that they could 
find in which $\Omega_{V}(t_0)$ is taken to be zero turn out to 
require negative $\Omega_{M}(t_0)$, viz. 
again negative $q(t)$ ($=\Omega_{M}(t)/2)$, and 
now negative values of $G\rho_M(t)$, i.e. effectively repulsive 
gravity. In addition, such a cosmology also has negative $k$. 
In passing the groups also note that a fit with no matter at all 
($\Omega_{M}(t)=0$, $\Omega_{V}(t)=0$), viz. the pure negative 
curvature case ($R(t)=(-k)^{1/2}ct+R(t=0)$, $q(t)=0$, 
$H(t_0)d_L=(1+z)log(1+z)$) is just at the allowed limit of their 
systematic errors. The fits to the data thus all point in the 
direction of the existence of some sort of cosmic repulsion, and 
indirectly suggest that it may even potentially be related to the 
spatial curvature of the universe.

While the above analysis shows that it is phenomenologically 
possible to find values of $\Omega_{M}(t_0)$ and $\Omega_{V}(t_0)$ 
for which standard gravity is able to fit the data, it is 
important to recognize that all of the inferred allowed 
values pose a severe theoretical challenge to the standard 
theory. Specifically, only one solution to the familiar flatness 
fine tuning problem of the standard cosmology has so far been 
seriously considered, viz. inflation, a solution which then 
requires $\Omega_{M}(t)+\Omega_{V}(t)=1$ to a very high degree of 
accuracy. However, in order to implement this condition, it is 
necessary to fix the value of the cosmological constant $\Lambda$, 
an as yet totally unresolved issue in standard gravity. Indeed, in 
standard gravity $\Omega_{V}(t_0)/\Omega_{M}(t_0)$ could be as 
large $10^{120}$. Because this is such an enormous number it has 
widely been believed that there would exist some as yet 
unidentified mechanism (such as some symmetry) which would then 
bring $\Omega_{V}(t)$ all the way down to zero. However, the 
ensuing $\Omega_{M}(t)=1$ cosmology now appears to have been 
excluded by the new high $z$ data. While inflation could still 
be saved by having non-zero $\Omega_{V}(t)=1-
\Omega_{M}(t)$, to 
achieve this would appear to require a fine tuning for
$\Omega_{V}(t)$ which would then be up to 60 orders of magnitude
more severe than the flatness one for which inflation was 
introduced in the first place.\footnote{In passing we note that 
the very 
introduction of such a non-zero $\Omega_{V}(t)$ would then change 
the expansion rate of the universe to 
$R(t)=(A_M/\Lambda)^{1/3}sinh^{2/3}\theta(t)$, to then potentially 
affect standard primordial nucleosynthesis calculations, and
especially so if $\Omega_{V}(t)\simeq 1$, a situation where $R(t)$ 
would then have to be growing at a close to exponential rate.} 
Thus a non-vanishing $\Lambda$ 
only serves to make the cosmological constant problem even more 
severe than before, with these 
new data essentially obliging the community to now have to deal 
with the cosmological constant problem head on.

Faced with the severity of this longstanding issue, a few years 
ago Mannheim \cite{Mannheim1990} suggested that the problem lay
not with some as yet unspecified dynamics, but rather with 
the very starting point of standard gravity, viz. the use of the 
Einstein-Hilbert action in the first place. Specifically, he 
noted that there is currently no known principle 
which would actually select out the Einstein-Hilbert action as the 
unique one for gravity (and thus none to simultaneously constrain 
the cosmological constant as well), and that one could just as 
well 
consider other equally covariant pure metric theories of gravity 
based on higher order derivative actions instead. Motivated by the 
fact that the three other fundamental interactions all possess 
actions with dimensionless couplings and dynamically induced 
masses, he proposed (as had others before him - dating back as far
as Weyl in fact) that gravity be endowed with an additional
underlying conformal invariance (viz. invariance under 
$g_{\mu \nu} (x) \rightarrow \Omega (x) g_{\mu \nu} (x)$), an
invariance which then yields 
$I_W=-\alpha \int d^4x (-g)^{1/2} C_{\lambda\mu\nu\kappa} 
C^{\lambda\mu\nu\kappa}$ (where $ C_{\lambda\mu\nu\kappa}$ is 
the conformal Weyl tensor and $\alpha$ is purely dimensionless) 
as its unique gravitational action. Conformal symmetry thus
emerges as a symmetry that  
naturally sets any fundamental cosmological constant term to zero. 
Moreover, Mannheim also 
noted \cite{Mannheim1990,Mannheim1992,Mannheim1995} that
because of the tracelessness of the conformal energy-momentum
tensor any cosmological term which might be induced by the 
spontaneous breaking of the conformal symmetry would have to be of 
order the energy density of all the other terms in 
$T^{\mu \nu}$, so that $\Omega_{V}(t)$ would then nicely be 
precisely of order the energy density associated with 
$\Omega_{M}(t)$ and/or of order that associated with the 
gravitational field itself, to thus make it possible to 
naturally yield an $\Omega_{V}(t)$ term of order $\Omega_{M}(t)$. 

To illustrate this point explicitly it is convenient to consider 
as matter action the typical case of fermionic matter fields 
Yukawa coupled to scale breaking scalars. For them, the most 
general covariant, conformal invariant matter action $I_M$ takes 
the form
\begin{equation}
I_M=-\hbar \int d^4x(-g)^{1/2}[ S^\mu S_\mu/2+ \lambda S^4-
 S^2R^\mu_{\phantom{\mu}\mu}/12+
i \bar{\psi}\gamma^{\mu}(x)(\partial_\mu+\Gamma_\mu(x))\psi -
gS\bar{\psi}\psi]   
\label{(2)}
\end{equation}
where $\Gamma_\mu(x)$ is the fermion spin connection and $g$ and 
$\lambda$ are dimensionless coupling constants. When the scalar 
field $S(x)$ in $I_M$ acquires a non-vanishing vacuum expectation 
value (which we are free to set equal to a spacetime independent 
constant $S$ because of the conformal invariance), the fermion 
then obeys the curved space Dirac equation 
$i\hbar \gamma^{\mu}(x)[\partial_{\mu} +\Gamma_\mu(x)]\psi = 
\hbar g S \psi$ and acquires a mass $\hbar gS$. For macroscopic 
purposes we note that the incoherent averaging of the fermion 
kinetic energy operator 
$i\hbar \bar{\psi}\gamma^{\mu}(x)(\partial_\mu+\Gamma_\mu(x))\psi$ 
over all the occupied positive frequency modes of this Dirac 
equation leads us \cite{Mannheim1992} to a standard 
kinematic perfect fluid of these fermions 
with energy-momentum tensor $T^{\mu \nu}_{M}=(\rho_{M}
+p_{M})U^\mu U^\nu+p_{M}g^{\mu\nu}$; while the averaging of the 
total energy-momentum tensor and of the scalar field equation of 
motion associated with Eq. (\ref{(2)}) respectively lead us to the 
equations    
\begin{equation}
T^{\mu\nu}=T^{\mu \nu}_{M}
-\hbar S^2(R^{\mu\nu}-
g^{\mu\nu}R^\alpha_{\phantom{\alpha}\alpha}/2)/6            
-g^{\mu\nu}\hbar\lambda S^4
\label{(3)}
\end{equation}
\begin{equation}
\hbar S^2R^\alpha_{\phantom{\alpha}\alpha}-
24\hbar \lambda S^4+6(3p_{M}-\rho_{M})=0
\label{(4)}
\end{equation}
\noindent
The covariant conservation of the total $T^{\mu\nu}$ thus
enforces that of the kinematic $T^{\mu \nu}_{M}$ as well, so that
$T^{\mu \nu}_{M}$ behaves exactly the same way in conformal 
gravity as it does in standard gravity.   
Because of the tracelessness of the total conformal $T^{\mu\nu}$,
Eqs. 
(\ref{(3)}) and (\ref{(4)}) immediately entail that the induced 
cosmological 
$\hbar\lambda S^4=\Lambda$ term is precisely of order $\rho_{M}$ 
and/or of order $\hbar S^2R^\alpha_{\phantom{\alpha}\alpha}$, 
with its explicit numerical value self-consistently adjusting 
itself to however many fermionic modes are occupied in 
$T^{\mu\nu}_{M}$ in whatever is the appropriate geometry. It is 
in this way then that the cosmological constant is kept under 
control in conformal gravity, with it being precisely the lack of 
any such self-consistency which leaves the cosmological 
constant problem unresolved in standard gravity. As we 
shall now see, it is precisely this very same conformal 
self-consistency which will 
lead us naturally to the negative curvature dominated 
universe suggested by the new high $z$ data.

To this end we note that the appropriate geometry needed for
Eqs. (\ref{(3)}) and (\ref{(4)}) is given by the  
conformal gravity field equations $(-g)^{-1/2}\delta I_W/\delta 
g_{\mu \nu}=-2\alpha W^{\mu\nu}=-T^{\mu\nu}/2$. Since $W^{\mu\nu}$
vanishes identically in a Robertson-Walker geometry (this metric 
being conformal to flat), in cosmology the conformal gravity 
equations reduce to $T^{\mu\nu}=0$, to yield 
\begin{equation}
\dot{R}^2(t) +kc^2 =
-3\dot{R}^2(t)(\Omega_{M}(t)+
\Omega_{\Lambda}(t))/ 4 \pi S^2 L_{PL}^2
\equiv \dot{R}^2(t)\bar{\Omega}_{M}(t)+
\dot{R}^2(t)\bar{\Omega}_{\Lambda}(t)
\label{(5)}
\end{equation}
written here in a particular convenient and suggestive form which 
involves $L_{PL}=(\hbar G/c^3)^{1/2}$ and the previously 
introduced $\Omega_{M}(t)$, though the ratio 
$\Omega_{M}(t)/ L_{PL}^2$ is of course independent of $G$. As we
see, Eq. (\ref{(5)}) only differs from the analogous standard 
model Eq. (\ref{(1)}) in one regard, namely that the 
quantity $-\hbar S^2 /12$ has replaced $c^3/16 \pi G$. From the 
point of view of the standard model, Eq. (\ref{(5)}) would 
have been obtained in standard gravity if standard gravity were 
repulsive rather than attractive, with the back reaction of the 
scalar field on the geometry in conformal gravity thus acting like 
an induced effective repulsive rather than attractive 
gravitational term in the conformal case, to thus nicely generate
what would have been regarded in the standard model as a
negative $G\rho_{M}$ term (an apparently allowed solution to the 
analysis of the recent high $z$ data given above). Because of 
this crucial change in sign, the $\dot{R}^2 $ and 
$\dot{R}^2\bar{\Omega}_{M}(t)=-
3\dot{R}^2\Omega_{M}(t)/ 4 \pi S^2 L_{PL}^2$ terms are not 
required to cancel each other so that the fine tuning flatness 
problem present in the standard model is simply not encountered in 
the conformal case \cite{Mannheim1992}. Moreover, unless the 
coefficient $\lambda$ is overwhelmingly negative, Eq. 
(\ref{(5)}) can only have solutions at all if $k$ is negative, 
and we are thus led naturally to a topologically open, negatively 
spatially curved universe. (Essentially the only way the 
geometry can cancel 
the positive energy density of ordinary matter and maintain 
$T^{\mu \nu}=0$ is if the gravitational field itself contains the 
negative energy associated with negative spatial curvature.) 
Conformal gravity thus nicely generates
the negative $k$ solution suggested by the analysis of the 
recent high $z$ data given above.

Before discussing further the cosmological implications of 
Eq. (\ref{(5)}), since we do have an effective repulsive
Einstein tensor in Eq. (\ref{(3)}), we need to see whether the 
conformal theory nonetheless has a good non-relativistic 
Newtonian limit. To this end we note that Mannheim and 
Kazanas \cite{MannheimandKazanas1989,MannheimandKazanas1994} found 
that for a static, spherically symmetric source, the equations
of conformal gravity reduce exactly (and without any 
approximation whatsoever) to $\nabla^4 B(r)=f(r)$ where 
$B(r)=-g_{00}(r)$ and $f(r) ={3(T^0_{{\phantom 0} 0} - 
T^r_{{\phantom r} r})/4\alpha B(r)}$; with the 
solution exterior to a source such as a star then taking the 
generic form $ds^2=B(r)c^2dt^2- dr^2/B(r)-r^2d\Omega$ where 
$B(r)=1-2\beta^{*}/r+\gamma^{*} r$ (see also Riegert 
\cite{Riegert1984}), and with the coefficients  
obeying  $\beta^{*}=\int_0^{R_0}drr^4f(r)/6$ and $\gamma^{*}=
-\int_0^{R_0}drr^2f(r)/2$ for a star of radius $R_0$. We thus see 
that for small enough $\gamma^{*}$ all the classic 
Schwarzschild metric tests of General Relativity can also be met 
in conformal gravity (so that the Einstein-Hilbert action turns 
out to only be sufficient to yield Schwarzschild but not
necessary), with the obtained conformal gravity Newtonian 
potential term 
being universally attractive for an appropriate sign of the 
fundamental coupling constant $\alpha$. 
Conformal gravity thus demotes $G$ from fundamental status (to 
incidentally thereby demote the Planck length $L_{PL}$ from 
fundamental status and thus decouple it from quantum gravity 
fluctuations), with the sign of the coefficient $\beta^{*}$ 
bearing no relation at 
all to the sign of the induced Einstein tensor term obtained in 
Eq. (\ref{(3)}). Solar system gravity is thus 
seen to not be a good guide as to whether gravity is attractive or 
not on altogether different distance scales, with the difficulties 
that standard gravity faces cosmologically perhaps being 
ascribable to its insistence that gravity actually be attractive 
(and of strength $G$) on those scales too.

Given the  uncovering of the above linear $\gamma^{*}r$ term
we immediately have to ask whether there is any evidence which
might support its possible existence. To this end 
Mannheim \cite{Mannheim1996,Mannheim1997} noted an interesting, 
purely phenomenological fact regarding galactic rotation curves. 
Specifically, he found that the centripetal accelerations of
the furthest data points in a large class of well studied spiral 
galaxies all obey the universal relation
\begin{equation} 
(v^2/c^2R)_{last}=\gamma_0/2+\gamma^{*}N^{*}/2 
+\beta^{*}N^{*}/R^2
\label{(5a)}
\end{equation} 
where the two new universal constants 
$\gamma_0$ and $\gamma^{*}$ take numerical 
values $3.06\times 10^{-30}$ cm$^{-1}$ (a value strikingly close 
to the inverse of the Hubble radius) and 
$5.42\times 10^{-41}$ cm$^{-1}$ respectively. This purely 
empirical formula (which dark matter halo models are therefore now
obliged to reproduce) then reveals the existence of not one but 
two linear potentials, with only one ($\gamma^{*}N^{*}c^2r/2$) 
being
associated with the material inside the galaxy. As regards the
galaxy independent term associated with $\gamma_0$, Mannheim went 
on to show that it was generated by the material outside of each 
galaxy, i.e. by the rest of the universe. (Once the potential is 
not of the $1/r$ form exterior matter sources can never be 
ignored,
and especially so if their potentials actually grow with 
distance.) Moreover, since the general coordinate 
transformation $r=\rho/(1-\gamma_0 \rho/4)^2$,  
$t = \int d\tau / R(\tau)$ effects the metric transformation
\begin{eqnarray}
(1+\gamma_0 r)c^2dt^2-{dr^2 \over (1+\gamma_0 r)}-r^2d\Omega 
\rightarrow
\nonumber \\
{(1+\rho\gamma_0/4)^2 \over R^2(\tau)(1-\rho\gamma_0/4)^2}
\left(c^2 d\tau^2 - {R^2(\tau) (d\rho^2 + \rho^2 d\Omega)\over
(1-\rho^2\gamma_0^2/16)^2}
 \right) 
\label{(5b)}
\end{eqnarray}
we see that a static linear potential is conformally and 
coordinate equivalent to none other than a Robertson-Walker
metric with the 
explicitly negative curvature given by 
$k=-(\gamma_0/2)^2=-2.3 \times 10^{-60}$ cm$^{-2}$. 
Thus in its rest frame each comoving galaxy sees the entire 
cosmological Hubble flow acting as none other than a universal 
linear $\gamma_0 c^2r/2$ potential term generated by the spatial 
curvature of 
the universe. With the use of these two linear potentials 
conformal gravity was then
able \cite{Mannheim1996,Mannheim1997} to provide for a complete, 
parameter free accounting of rotation curve systematics without 
the need to invoke galactic dark matter. We thus identify an 
explicit cosmological imprint on galactic rotation curves, 
recognize that it was its neglect that may have led to the need 
for galactic dark matter, and, for our purposes here, confirm from 
galactic data that the cosmological curvature $k$ is indeed 
negative, just as desired for the new high $z$ data. 

Having now established some empirical evidence that $k$ is indeed 
negative, we return to an examination of the cosmology based on 
Eq. (\ref{(5)}). For the (simpler to treat) radiation era where
$\rho_{M}=A/R^4=\sigma T^4$, exact $k$ negative solutions can be 
found \cite{Mannheim1992,Mannheim1995}, with these solutions 
being sensitive only to the sign of $\lambda$. For $\lambda>0$, 
$\lambda=0$, and $\lambda<0$ they are respectively given as  
\begin{eqnarray}
R^2(t)=R_{min}^2+(R_{max}^2-R_{min}^2)sin ^2 (\pi t / \tau)~~,~~
R^2(t)=-2A/k\hbar c S^2-kc^2t^2~~,~~
\nonumber \\
R^2(t)=R_{min}^2+|R_{max}^2-R_{min}^2|sinh ^2 (\pi t / \tau)
\label{(6)}
\end{eqnarray}
where $\tau=\pi/cS(2|\lambda|)^{1/2}$ and where 
$(R_{max}^2,~R_{min}^2)=
(-k/4S^2\lambda)(1 \pm (1-16A \lambda /k^2\hbar c)^{1/2})$. 
In these solutions $\Omega_M(t)$ takes the values
\begin{eqnarray}
\Omega_{M}(t,\lambda>0)=\Omega_{min}/sin^2(2\pi t/\tau)~~,~~
\Omega_{M}(t,\lambda=0)=8\pi GA/3c^6k^2t^2~~,
\nonumber \\
\Omega_{M}(t,\lambda<0)=\Omega_{min}/sinh^2(2\pi t/\tau)
\label{(7)}
\end{eqnarray} 
where 
$\Omega_{min}=16\pi G A/3c^4|\lambda| S^2(R_{max}^2-R_{min}^2)^2$,
a quantity which can readily be set to be less than one for a 
continuous range of parameters. Since each $\Omega_{M}(t,\lambda)$ 
is infinite at $t=0$ (in sharp contrast to the standard model 
where $\Omega_{M}(t=0)=1$) and since they go to either zero or 
$\Omega_{min}$, each $\Omega_{M}(t,\lambda)$ has to pass through 
one at some time without any need for fine tuning, so that 
$\Omega_{M}(t_0)$ can readily be reasonably close to one today 
without needing to be identically equal to one, with the 
copious amounts of cosmological $\Omega_{M}=1$ dark matter thus 
not being needed in the conformal case.    
Whether or not the cosmology recollapses is seen to depend on the 
sign of $\lambda$ rather than on that of $k$ (the case in the 
standard model). Moreover, in each case there is a non-zero 
minimum radius (and thus a maximum temperature $T_{max}$), with 
the cosmology thus being singularity free
(precisely because it induces a repulsive gravitational component 
so that conformal gravity can protect itself from its own 
singularities - something of course not the case in standard 
gravity). The $\lambda>0$ cosmology is the only one
which recollapses, with a full cycle then taking a time $\tau$.
Since there are no singularities, this cycle is then free to 
repeat indefinitely.

All of the key features of these solutions can be exhibited 
directly in the $\lambda=0$ case, with Eq. (\ref{(6)}) then 
yielding
\begin{eqnarray}
T^2_{max}/T^2(t)=1/ (1-tH(t))=1-1/ q(t)
=1+4\pi L^2_{Pl}S^2/ 3\Omega_{M}(t)~~~,~~~
\nonumber \\
(-k)^{1/2}c\int_0^t dt /R(t)=log[T_{max}/T+
(T_{max}^2/T^2-1)^{1/2}]
\label{Eq. (7)}
\end{eqnarray}
\noindent
Since $\Omega_{M}(t_0)$ is of order one today 
the scale parameter $S$ must be at least as big as 
$10^{10}L_{PL}^{-1}$ if the early universe is to have a 
temperature of at least $10^{10}$ degrees. With this one 
(natural) condition it then follows in the $\lambda=0$ case that
the current value of the deceleration parameter is zero and that 
the age of the universe is $t_0=1/H(t_0)$, an age which nicely
exceeds current globular cluster age limits. Further, in 
passing, we see that 
the horizon size is altogether greater than one at recombination, 
to thus yield a naturally causally connected cosmology. 

Because of the huge value needed for $S$ (i.e. because $L_{PL}$ 
does not set the scale for conformal cosmology) Eq. (\ref{(5)}) is 
found \cite {Mannheim1995} to be realized in a very interesting 
way, with the $\dot{R}^2(t)$ and 
$\dot{R}^2(t)\bar{\Omega}_{M}(t)= 
-3\dot{R}^2(t)\Omega_{M}(t)/ 4 \pi S^2 L_{PL}^2$ terms 
being found to have radically differing time behaviors, even as  
the quantity $\dot{R}^2(t)(1-\bar{\Omega}_{M}(t))$ must remain
equal (in the $\lambda=0$ case) 
to the constant value $-c^2k$. Specifically, as can be seen from 
Eq. (\ref{(7)}), the quantity $\bar{\Omega}_{M}(t)$ can only 
be large in the early universe (where $\rho_{M}$ sets the scale 
for $R(t=0)$ in Eq. (\ref{(6)})) with it having to fall to 
a value of order at least $10^{-20}$ today, thus causing the 
$\dot{R}^2(t_0)$ term to be completely dominant over 
$\dot{R}^2(t_0)\bar{\Omega}_{M}(t_0)$ today. 
Thus even while $\Omega_{M}(t_0)$ is nicely of order one 
today in conformal gravity, the specific conformal gravity source 
$\bar{\Omega}_{M}(t_0)$ itself is completely negligible 
today, with the late time (but not the early time) $\lambda=0$ 
conformal cosmology then being curvature dominated. Moreover, this 
same late
time decoupling of $\bar{\Omega}_{M}(t_0)$ also holds for both
the $\lambda \neq 0$ case and the matter $\rho_M=A_M/R^3$ era as 
well \cite {Mannheim1995}. Late time conformal cosmology is thus
always curvature and/or $\lambda$ dominated, and thus even while 
$\Omega_{M}(t_0)$ is non-zero, nonetheless the conformal gravity 
evolution equation Eq. (\ref{(5)}) reduces at late times to
$\dot{R}^2(t) +kc^2 =\dot{R}^2(t)\bar{\Omega}_{\Lambda}(t)$. Thus
the late time conformal cosmology evolves exactly the same way
as the standard model Eq. (\ref{(1)}) would evolve if the standard 
model $\Omega_{M}(t_0)$ were to be set to zero. Thus the 
acceptable $\Omega_{M}(t_0)=0$, $\Omega_{V}(t_0)\geq 0$ fits 
considered by the observers in their standard model analysis of 
the new high $z$ data thus immediately apply to late time
conformal cosmology case as well, with the attendant positive 
fitted values reported for $\Omega_{V}(t_0)$ entailing that 
the conformal cosmology parameter $\lambda$ would then be 
negative.
Conformal cosmology thus provides a natural (i.e. non fine tuned) 
explanation for a curvature and/or $\lambda$ dominated late 
universe, a cosmology which naturally produces a cosmic repulsion
and which appears to be fully compatible with currently available 
high $z$ data.

As we have thus seen, rather than being $\rho_{M}$ dominated, 
conformal cosmology actually becomes curvature and/or 
$\lambda$ dominated at late times. While this curvature 
dominated cosmology is thus seen to be able to address some 
outstanding puzzles of the $\rho_{M}$ dominated standard model, 
it is not itself yet completely free of problems, since it seems 
to be able to only produce substantial amounts of primordial
helium and appears to be incapable (at least in the studied small 
$\lambda$ limit) of reproducing other light 
element abundances \cite{Knox1993,Elizondo1994}. Whether this is 
simply a property of using just the simple cosmology based on 
Eq. (\ref{(5)}) or its $\lambda=0$ limit (the now suggested large 
negative
$\lambda$ region still remains to be explored), and/or whether it 
could be resolved in more detailed dynamical conformal models 
remains to be addressed. However, independent of one's views 
regarding conformal gravity itself, the study we have presented 
here does provide a new and quite general way to approach 
cosmological issues, showing that once one is prepared to depart 
from the Einstein Equations, new options then become available. 
Additionally, our study has enabled us to explicitly determine
the global topology of the universe, something which years of work 
in standard gravity has yet to accomplish. Since we have also 
shown that the cosmological constant problem, the flatness 
problem, the 
horizon problem, the dark matter problem, the universe age 
problem, and now the recent high $z$ cosmic repulsion problem are 
not in fact generic to cosmology, their very 
existence may thus even be a signal that the extrapolation of 
standard gravity from the solar system to cosmology might be a lot  
less reliable than is commonly believed. 
The author would like to thank Dr. B. Schaefer for helpful
comments. This work has been supported in part by the 
Department of Energy 
under grant No. DE-FG02-92ER40716.00.

\end{document}